\documentclass[runningheads]{svmult}

\usepackage{makeidx}   
\usepackage{graphicx}  
\usepackage{subeqnar}  
\usepackage{multicol}  
\usepackage{physprbb}  
\makeindex             

\begin{document}
\title*{Epilogue}
\toctitle{Epilogue}
\titlerunning{Epilogue}
\author{Roger Blandford}
\authorrunning{Roger Blandford}
\institute{130-33 Caltech, Pasadena, CA 91125}
\maketitle
\begin{abstract}
There are now several types of relativistic flows in 
astrophysical settings. The foremost examples are
jets and disks orbiting spinning 
black holes, pulsar winds and gamma ray bursts. 
As discussed at this meeting, these flows exhibit unusual 
kinematic and dynamical properties, 
that distinguish them from non-relativistic flows.
It is possible that all collimated outflows are
essentially hydromagnetic or electromagnetic. 
Future study of relativistic flows will rely heavily on numerical
experiments. Model relativistic flows
provide a basis for carrying out secondary studies of the underlying plasma 
physics, particle acceleration, magnetic field amplification and
the emission and transfer of radiation, particularly
at shock fronts. Some current opportunities in 
observation, phenomenology and theory are briefly suggested.  
\end{abstract}
\section{Relativistic Flows in Astrophysics}
Although the initial forays into relativistic gas dynamics 
\cite{tau48} and relativistic MHD \cite{lic67} were mainly 
stimulated by applied mathematical curiosity, astronomical observations
now provide abundant incentive to consider relativistic flows in detail.
In this brief summary, I shall review these developments in rough, 
historical order, emphasise some contemporary problems and
suggest some future directions. I refer to the many excellent 
presentations at this workshop for`detailed discussions and more 
extensive bibliographies.
\subsection{AGN}
Ever since the pioneering, VLBI observations in the early 1970's 
\cite{coh71}, \cite{whi71} we have 
known that compact, extragalactic radio sources can expand with space 
velocities within one percent of the speed of light so as to create
a strong ``superluminal'' illusion \cite{ree66}. The 
emitting features, in variable, compact radio sources,
were soon associated with external and internal shock waves and the  
jet-like, as opposed, to spherical character of the flows
was established. Observations of jets at optical, X-ray and,
especially $\gamma$-ray energies followed and much effort is currently being
applied to understanding the details of how different types
of jet emit throughout the electromagnetic
spectrum. The general picture that has emerged 
[Agudo, Celotti, Close, Georganopoulos, Kovalev, Mastichiadis, Matzac]
is that relativistic 
electrons are accelerated, mainly at shock fronts, and that they emit 
synchrotron radiation at low frequency and inverse Compton emission at high
frequency, with the former providing the soft photons for the latter
in low power
sources and the disk ultraviolet emission supplying the soft 
photons in high power sources.

Although early, relativistic jet models were essentially 
gas dynamical, with the magnetic
field evolving passively, it is now widely believed that jets are 
collimated by large-scale magnetic field [Casse, Ferreira, Nitta]  
with a disk angular velocity vector determining the jet axis.
However the details are controversial. Most models
of jet formation and collimation that have been published to date are 
non-relativistic, analytic MHD solutions where the intrinsic 
anisotropy of the Maxwell stress tensor is ultimately responsible for the
collimation. This is reasonable because, as it is
generally reasoned, what is being described is a collimating, magnetic 
sleeve that confines an ultrarelativistic flow 
that emanates from the black hole and the inner disk.
However, some authors have argued that the magnetic field is always
primarily poloidal and any toroidal field that is generated by the rotation
will quickly vanish through reconnection. At the other extreme it has been 
proposed that rotation dominates and the magnetic field lines
behave like a coiled spring pushing the jets out along the spin axis. 
Both components are relevant to centrifugal models where the inertia
of the outflowing plasma plays a crucial role. Finally, there are 
models where there is no long range order to the field and it is the local 
anisotropy associated with loops of magnetic flux that create the collimation.
The stability of most of these models, particularly to non-axisymmetric modes,
is only recently coming under scrutiny [Lery].

Yet another debate, and this is probably the central phenomenological
question in interpreting relativistic flows, is ``What is the working
substance?''. [Markowith, Yamasaki]. There is 
far too large a radiation density around
the inner disk for the flows to start life as just pairs and there has to be 
some other carrier of momentum. One possibility, considered in some early 
models was the radiation itself. However, 
elementary kinematic considerations make it quite unlikely that
large Lorentz factors can be achieved. Protons, whose
radiative efficiency is very low, could be responsible, though it is 
very hard to see how they could be accelerated efficiently in a beam.
In my view, the most reasaonable candidate is electromagnetic field. 
Note, that I am only suggesting that jets start off in a predominately
electromagnetic form. A quite likely sequence is that, at some
finite distance from the black hole where annihilation cannot
keep up with production, jets metamorphose into 
a pair plasma. This may be where the observed $\gamma$-rays are produced.
At a yet greater distance, these jets should ultimately interact strongly
with their surroundings as they become radio sources and decelerate.
Presumably, 
when the jet is powerful, the outflow can remain relativistic and we have
an FR2 source; when the jet is weak and decelerates to a subsonic 
speed, an FR1 source is formed. At this point, we are dealing with 
reasonably well-resolved extended radio sources and now have a much
better understanding of the physical conditions in the surrounding gas.
This should enable us to make more confident descriptions of radio
source evolution and more quantitative estimates of the 
total jet powers [Blundell, Manolakou, Polatidis], 
both in individual sources (where they can exceed
the bolometric power of the AGN) and collectively as a
contribution to the luminosity density of the universe. 

Although jets are certainly the most dramatic, relativistic 
flows associated with AGN, gas in the accretion disk also moves with 
mildly relativistic speed and general as well as special relativity must
be invoked to account for the X-ray line spectroscopy - a quite dramatic
vindication of the black hole model. This is  possible evidence that the 
holes are spinning, because the prominent red wings 
that are sometimes observed should only be formed 
if the disk can approach the horizon and this onloy happens for a 
geometrically thin disks orbiting a rapidly spinning hole
in a prograde sense \cite{tan95}.
\subsection{Pulsar Winds}
Pulsars were discovered soon after there was evidence for relativistic 
effects in AGN and it was quickly realized that they should 
also be (relativistic) electromagnetic objects \cite{pac67}\cite{gol68}. 
Originally, the field was thought to be that of a rotating, vacuum, 
magnetic dipole, though it was soon realized that the magnetosphere
had to contain plasma which would at least seriously modify the 
electromagnetic field and might have dynamical importance \cite{gol69}.
A similar metamorphosis of energy from mechanical, through electromagnetic,
pairs and ions is envisaged though the details of how and where these
transformations occur are no less controversial than they are with AGN jets
despite much valiant, theoretical effort.

Important Chandra observations of the Crab Nebula, and a few other 
plerions, have 
demonstrated that the presumed spherical winds actually
exhibit ``jets'', giving the lie to the assertion
that disks are necessary for jet formation. (Actually, they also appear 
to possess features that look like ``disks'', though these are probably 
equatorial current sheets, like those found to be associated 
with the solar wind.) 
\subsection{Galactic Superluminal Sources}
The association of black hole accretion disks with AGN led to 
the (morphological)
expectation that binary X-ray sources (where the direct evidence for disks 
was stronger than in AGN) should also produce jets. 
The early evidence (eg in Sco-X1) was 
confused, but with the discovery of the jets in SS433 \cite{mar79} [Rowell], 
(where the jet velocity 
and its variation could be accurately measured) the matter was settled.
However, here, and in all other known examples to date [Fender], 
the outflow speed is only mildly relativistic in contrast 
to what has been found with the 
AGN jets. Indeed, there is no dynamical or kinematical
objection to the jets
being created by radiation pressure and as these sources are operating 
quite close to the Eddington limit, radiation cannot be ignored.
These inner disks, like those associated with AGN accreting close to the
Eddington rate, comprise radiation-dominated gas.

The Galactic superluminal sources \cite{mir94}, which 
are believed to contain black holes, also exhibit ``Quasiperiodic
Oscillations'', or QPOs, analogous to the oscillations that have been reported
from neutron star systems. The modulation is generally
associated with standing modes in 
the relativistic accretion disk. However, the disk can only be the clock;
the X-ray emission is so hard that it must actually originate in the corona.
This is additional, circumstantial 
evidence for a strong magnetic coupling between the disk
and its surroundings and, in principle, a strong diagnostic of general
relativistic disk flow. 
\subsection{Gamma-Ray Bursts}
Although early discussions of the nature of GRBs clearly recognized 
the implications if they were at cosmological distances \cite{rud75}, 
it wasn't until the BATSE catalog was produced that 
it became clear that this was probably the case and that
GRBs probably expanded even faster than AGN jets \cite{pac86}.
The inference was verified by the measurement of afterglow redshifts 
and the discovery of radio scintillation[Downes, Kobayashi, Galama, Sari]. 
It now seems to be generally
accepted that bulk Lorentz factors, variously estimated as lying in 
the range $100<\Gamma<1000$, are required to avoid pair production 
by the escaping high energy $\gamma$-rays.  
The gamma ray burst itself is most commonly associated with the dissipation
of internal shocks that form in the expanding fireball and reflect
variation in the source over a relatively long timescale $\sim100$~s for 
the better studied ``long'' bursts. The afterglow, which can be traced
for over  year in some cases, is associated with a blast wave, initially
ultrarelativistic, formed by the swept up interstellar medium.
The evidence that this flow is non-spherical, ie that GRBs are also jets 
-- the observation of achromatic spectral breaks and 
a desire to limit the explosion energy -- is improving but is not yet
decisive. 

The study the dynamics and radiative properties of 
afterglows has partly recapitulated 
the study of AGN jets, although there is now an impressively detailed
phenomenological description of comprehensive observations of over
20 bursts throughout the electromagnetic spectrum.
Although there is circumstantial evidence that the long bursts are associated 
with star-forming regions in galaxies, the nature of the sources is 
still unclear. we know less about the short bursts, though the 
soft gamma repeaters are probably associated with magnetars.  

In most contemporary models of the non-repeating bursts, a black hole
is either formed or augmented.
Many of these models, specifically the collapsar models, raise fundamental
questions of relativistic gas dynamics, including the 
question of whether or not we are dealing with a fluid at all! 
The afterglows raise the same questions that came up with AGN 
concerning particle acceleration and field amplification.
particles accelerated and how is field amplified?''. From an astronomical
perspective, we also want to understand the place of GRBs in 
the scheme of advanced stellar evolution and supernova explosions
as well as their potential 
as sources of neutrinos and gravitational radiation as well as 
their ``environmental impact''.
\subsection{Other Relativistic Flows}
There are several other types of mildly relativistic flow that have been 
considered in astrophysics including accretion onto neutron stars, broad
absorption line quasar outflows and, most fundamentally of all,
early universe cosmology which is 
also an exercise in relativistic gas dynamics with a transition from 
radiation-dominance to gas-dominance, just like that in accretion disks!
\section{Relativistic Flows}
\subsection{Gas Dynamics}
One of many satisfying features of both special and general relativity is 
how harmoniously they accommodate gas dynamics. The relativistic
formalism emphasizes symmetry and conservation laws
in a manner that is sometimes lost in the more engineering-oriented 
development of the non-relativistic subject. 
Gravity can be ignored for application to jets, winds and GRBS 
and the governing
equations, derivable from setting the divergence of the mass 
particle current vector and the stress-energy tensor to zero, express
the conservation of mass, momentum and energy. This leads to  
counterparts of familiar non-relativistic descriptions for ID flow, shock 
discontinuities and so forth. In many analyses of relativistic flows, 
the fluid is often taken
to be isotropic and ultrarelativistic, that is to say the pressure is 
dominated by radiation or high energy leptons 
with an internal sound (proper) speed
of $2^{-1/2}c$. In this case the effective Mach number
is $M=2^{1/2}u$ where $u$ is the proper bulk speed. In other applications,
non-relativistic protons are also present and reduce the sound speed.

However, there are some serious worries
as to how complete a description this really is in many of the environments
where these results are applied. For example, in a pair creation region,
mass will not be conserved. Furthermore, momentum and energy
will not be conserved 
in the presence of inverse Compton scattering. 
Another worry is that shear stress
is usually ignored when dealing with jets
while it is seen as an intrinsic part of another 
common astrophysical shear flow, the accretion disk. 
When we make 1D jet models we are
implicitly assuming that jets are enclosed by narrow, turbulent, 
boundary layers
that do not spread so that Mach numbers can attain large values
-- over 300 in some collapsar models -- and the ratio of the 
bulk kinetic energy to the internal energy exceeds $\sim M^2$. This 
is supposed to happen naturally with essentially no noise and internal
dissipation reconverting the bulk energy to internal energy. 
An aerodynamicist would think this strange!

High Mach number jets have some unusual properties (both 
non-relativistically and relativistically). If the fluid starts
from a subsonic chamber, where it is all in causal contact, and 
accelerates through a pair of nozzles to form ``twin exhausts''
with Mach number much larger than the reciprocal of the 
jet opening angle $\theta^{-1}$, (as is thought to happen
in GRBs), then the different parts of the jet
flow will fall out of causal contact. Now, in the case of a GRB,
the jet is likely to be preceded by a relativistic
blast wave propagating into the surrounding medium. Initially the different 
elements of this blast wave will also be out of causal contact. 
However, as the blast wave  decelerates, $M$ will
fall to $\theta^{-1}$ and transverse causal contact will be 
re-established. This ``hello--goodbye--hello'' behavior is thought to be 
responsible for the achromatic breaks in the afterglow emission and is
reminiscent of inflationary cosmology!

Ultrarelativistic flows do have some distinctive kinematic properties
which mostly derive from the fact that the 3-speed is limited to that
of light. This in turn, leads to strong Doppler-shifting and beaming
of the emitted radiation. These effects can be extremely large and can lead to 
insignificant parts of the source dominating what we observe   
\subsection{Passive Gravitational Field}
The next most complicated class of problems involves gas dynamical 
flows in the presence of a passive gravitational field.
A prime example is an accretion disk in orbit about a black hole.
There is now a lot of interest in solving these 
problems using the full machinery of general relativity.
For example, ``diskoseismological'' oscillation
modes have been calculated and their frequencies 
can be made to match QPO observations. 
\subsection{Active Gravitational Field}
Flows where the spacetime is dynamic are far harder to analyze and 
numerical methods are necessary. The most pressing examples are neutron star
-- neutron star/black hole models of GRBs. Simulations 
have been used to determine the timescales for coalescence 
and, for example, to show that neutrino emission is unlikely 
to be very important in driving the burst.  
\subsection{Magnetohydrodynamics}
As I have already remarked, most models of relativistic jets, plerions 
and disks are intrinsically magnetised. In particular, we now know how magnetic
field is amplified in a non-relativistic disk, through the magnetorotational 
instability and it is now generally agreed that
disk evolution is a magnetohydrodynamic problem.
Similarly, jet collimation is generally argued to be due to 
anisotropic  magnetic stress on the grounds that the maximum gas pressures
allowed by X-ray observational constraints are too small to effect
collimation. 

In non-relativistic astrophysical MHD [Sauty], it is commonly assumed that the 
electrical conductivity is infinite, implying that the electric
field vanishes in the centre of momentum frame. When this 
happens the flow evolves under a set of locally deterministic
equations which give the partial derivatives of the velocity, density
and the magnetic field with respect to time.
(There is an applied mathematical 
nicety involving the degeneracy of the signal speeds along the field.
which is probably inconsequential in practice.) Note that there is no need
for an equation to describe the temporal evolution of the current density
as this is given by the curl of the magnetic field. The charge density can be 
determined after the fact from gauss' law, if needed, but it has no dynamical
role in the non-relativistic limit. 

By contrast, when we try to do the same thing in relativistic MHD, 
some awkward questions are raised because we cannot ignore 
the displacement current and the charge density. Maxwell's
equations are evolutionary equations for $\rho,\vec E, \vec B$.
In the infinite conductivity limit, there
is no evolutionary equation for $\vec j$. Only when
we introduce a finite conductivity, so that the current 
density is given by some form of Ohm's law in the center of momentum
frame, do we fix the current locally. However, when the conductivity
is so large that there is insignificant electric potential difference 
along the magnetic field, the current flow must be determined
by what happens elsewhere, in regions where there is dissipation or by the
boundary conditions.  

There is a second possible problem with relativistic MHD that can have a 
large bearing on the outcome. Traditionally there are three modes
of wave propagation known as fast and slow magnetosonic 
propagation, together with the intermediate (or Alfv\'en) mode.
Now the slow mode is determined by the sound speed in the gas. This 
is traditionally taken to be isotropic and (if the gas has a 
high temperature) to be
$3^{-1/2}c$. However, in many flows, including those around black holes,
it is possible for the electrons to cool on a dynamical timescale. The 
particle distribution function may become highly anisotropic with repect 
to the ambient magnetic field. In this case the effective
sound velocity along the direction of the magnetic
field can become arbitrarily close to the speed of light. 
This, in turn affects the characteristics and has implications 
for the development of shock waves and the causal structure of relativistic
flows.  
\subsection{Force-free Electrodynamics}
A useful approximation for handling magnetised, relativistic
flows, that simplifies the calculation,
though does not remove the first of the above difficulties, 
is to adopt the relativistic force-free
approximation, namely that $\rho\vec E+\vec j\times\vec B=0$. This immediately
supplies a constitutive relation for the component of the current density
resolved perpendicular to the magnetic field, $\rho\vec E\times\vec B/B^2$.
The parallel component 
must be fixed by boundary conditions, just as in the non-relativistic 
case. In this approximation, which is surely good for field lines
which thread a black hole event horizon and quite possibly for 
pulsar winds, we dispense with the velocity all together. The
role of the plasma is to supply charge density and current. 
The invariant $\vec E\cdot\vec B$ vanishes and there will generally 
be ample charge to keep the invariant $B^2-E^2$ positive.
\subsection{Radiation-dominated Gas Flows}
Increasing attention is being paid to the dynamics of radiation-dominated
fluid. The prime example is the open universe. However, here the vorticity
and magnetic field are thought to be quite small. By contrast, relativistic 
accretion disks and optically thick
jets are quite likely to behave very differently. These
are, by definition, shear flows in which magnetic fields
grow on a dynamical timescale. However, the field may become
quite inhomogeneous in a manner which will facilitate radiative transfer.
In fact, photons can transfer momentum as well as energy under some 
circumstances.
\section{Numerical Simulations}
It has been made quite apparent here that the way forward is through numerical 
simulation [Aloy, Peitz]. Impressive advances have been 
reported in the testing 
and deployment of large three (and four) dimensional codes with and 
without magnetic field and improvements in speed and memory make quite
sophisticated investigations a practical proposition. The numerical study of 
radiation-dominated, relativistic magnetohydrodynamics is on the horizon
and promises the biggest surprises as we strive to develop some 
understanding of what really happens to gas
accreting onto a black hole. Even in simple flows,
breaking spherical or axisymmetry and going beyond self-similarity
is producing large changes in our outlook. In addition, the 
capability to tackle non-linear perturbations - the only sort that observers
can see - is crucial.  
 
This, is not to say that there is no further
role for analytical approaches. In some sense, they become more important.
This is because it is extremely hard to represent the results of 
multi-dimensional computations, graphically or verbally in a manner 
that allows one to divine general principles and 
predict what will happen in other flows. Having a simple description
of the most important features of a complex flow is immensely
valuable. It is also important to distinguish
numerical simulation, which aspires to reproduce an accurate 
representation of a flow from numerical
experiments. As with much experimental physics,
a laudable goal of numerical experiments is to get so much insight that
it is possible to replace them with a working model that can be used
as a subunit of a larger investigation. 

Understanding the flow is not the end of the matter. It is important
to use the fluid solution to provide a framework to discuss higher 
order features like the plasma 
physics, particle acceleration and radiative transfer. This is vital
if we want to interpret the diagnostic observations of
relativistic sources. Carrying out these
secondary studies relativistically is turning out to be
no less of a numerical challenge than computing the basic flows. 
\section{What Now?}
\subsection{Observation}
The observational prospects are good. 
On AGN jets, there is an opportinity for using polarimetric observations
and imaging to tell us if jets have an electromagnetic or a gas dynamical
origin and for understanding what factors determine the jet power. 
Increasingly detailed X-ray observations
of sources like M87, Cygnus A and Pictor A are 
providing excellent laboratories for
determining directly, where relativistic electrons 
are accelerated, what is the magnetic feld geometry 
and how the particles are transported. Temporal studies 
should clarify the extent to which jets can be thought of
as continuous flows or a sequence of explosive outbursts. 
The spectacularly detailed X-ray spectroscopy that is emerging from 
both XMM-Newton and Chandra will eventually be interepreted and should 
define the geometry of the gas flow near to black holes
and, especially, identify where the energy is dissipated. 

Our view of plerions has been considerably enhanced by their association
with soft gamma repeaters and magnetars. However there is still a 
lot of uncertainty in these identifications and an imperfect understanding 
of how a plerion changes in reponse to a burst. More coordinated
observations are needed.

One of the big observational challenges
in studying the Galactic superluminal sources is to find a ``microblazar'' 
- a high Lorentz factor jet pointed towards us. 
(It is possible that the recently-discovered, super-Eddington 
compact sources in nearby galaxies could be of this type.) If, conversely,
we can persuade ourselves that these objects do not exist, then
it will probably tell us something useful about jet formation. 

Turning to GRBs, HETE2 should identify some
short bursts and point to a physical origin for this subclass to complement
the tremendous observations of the long bursts that derive from 
Beppo-SAX and BATSE.
\subsection{Phenomenology}
The basic jet emission model involving shock fronts admits a simple, 
testable prediction. The kinematic velocity of the emitting feature
will differ from the velocity of the emitting gas, which can be measured
through the Doppler shift. It would  be very nice to demonstrate this in some
sources.

There is great need to discover the true laws of MHD which will describe how 
collisionless plasmas behave in practice on the largest scales as 
opposed to the applied mathematical idealisations on which 
we must rely at present. Undoubtedly
our best hope for understanding non-relativistic flows
lies with careful analysis of the superb observations
of the solar corona by the YOHKOH, SOHO and TRACE spacecrafts as well 
as the {\it in situ} observations of the solar wind and planetary 
magnetospheres. We need to understand how currents flow -- are 
they distributed
or filamentary - how shocks create entropy under a wide variety of 
defining conditions, how much energy is dissipated in reconnection
regions and in what form, how turbulent spectra develop and the connection
to magnetic amplification by dynamo action. We would like to understand 
solar flares as a prelude to understanding the energisation of an accretion
disk corona and the means of launching the solar wind which is surely relevant
to the formation of jets. 

The best laboratory that we have for the relativistic flows that are the 
subject of this meeting is surely the Crab Nebula. Here Chandra and HST 
observations are changing our view of the pulsar wind/jet, and its 
termination through a strong shock front. However we still do not 
have an accepted determination of its speed and composition. The notion
that magnetised flows generically collimate into jets even without a disk, 
if true, is of immediate relevance to GRBs. 

A related question is ``How much is the character of the jet 
dependent upon the central compact object?''. We are pretty confident 
that the Galactic superluminals are identified with black holes
but Sco X-1 (and possibly, SS433) may derive from 
neutron stars. In addition
the jets we observe best are formed by protostars, so a ``compact object''
is far from necessary. This is a good clue as to how relativistic 
jets are powered. Perhaps all that is required to make 
a jet large relative angular velocity (in units of the Keplerian value).  

Another research frontier
is the Ultra High Energy cosmic rays. There is a good chance that, as 
observations continue to improve over the next few years, we will be forced
to a phenomenological model of their origin which will surely involve
ultrarelativistic plasma physics.  
\subsection{Theory}
There is now a large backlog of unsolved problems in fundamental 
theory that must be solved before we can model astrophysical, relativistic 
flows with confidence. For many of these, as I have emphasized, the requisite
computational tools are becoming available. Perhaps the most pressing need,
with the largest, general implications, is to understand how 
magnetic fields evolve around black holes, in the radiation
or ion-dominated disks and in the magnetosphere above the hole.
We will almost certainly need to perform large scale 
numerical simulations and will want to use these to determine
the relative efficiencies for the release of energy by 
the hole and the disk and the connection to jets.
This problem will be very hard to solve,
though it is fairly 
well posed. In the case of an ion plasma we will want to use
plasma simulations to understand better the collisionless
coupling between the hot ions and the cool electrons

There are several additional, interesting formal
challenges which may turn out to be
relevant to interpreting the observations, such as making a 
theory for the radiative transfer of plasma waves in a curved spacetime,
developing a relativistic theory of reconnection and reworking the 
theory of the MRI for a relativistic disk.

As the diverse contributions to this lively workshop attest, these are only 
a few of the possibilities inherent in this young and exciting field.  
\section*{Acknowledgements}
I am indebted to Markos Georganopoulos, Axel Guthmann, Konstantina Manolakou
and Alexandre Markowith for their invitation to this meeting, their
generous hospitality and allowing me to fulfil a lifelong ambition
to visit Greece. Support under NASA grant 5-2837 is gratefully
acknowledged.
\vfill\eject


\begin{thebibliography}{}
\addcontentsline{toc}{section}{References}
\bibitem{coh71}Cohen, M. H. {\it et al} 1971 ApJ 170 207
\bibitem{gol68}Gold, T. Nature 218 731
\bibitem{gol69}Goldreich, P. \& Julian, W. H. 1969 ApJ 157 869
\bibitem{lic67}Lichnerowicz, A. 1967 Relativistic Hydrodynamics and 
    Magnetohydrodynmics New York:Benjamin
\bibitem{mar79}Margon, B. {\it et al} 1979 ApJ 230 L41
\bibitem{mir94}Mirabel, I. F. \& Rodriguez, L. F. 1994 Nature 371 46
\bibitem{pac67}Pacini, F. Nature 216 567
\bibitem{pac86}Paczy\'nski, B. 1986 ApJ 308 L51
\bibitem{ree66}Rees, M. J. 1966 Nature 211 468
\bibitem{rud75}Ruderman, M. 1975 Ann. N. Y. Acad. Sci. 262 164
\bibitem{tan95}Tanaka, Y. {\it et al} 1995 Nature 375 659
\bibitem{tau48}Taub, A. H. 1948 Phys. Rev. 74 328
\bibitem{whi71}Whitney, A. R. {\it et al} 1971 Science 173 225
\end{thebibliography}
\end{document}